\documentclass[12pt]{article}

\usepackage{indentfirst}
\usepackage{amsmath, amssymb, amscd, amsthm, amsfonts}
\usepackage{xr}
\usepackage{bm}
\usepackage{mathtools}
\usepackage{graphicx}
\usepackage{enumerate}
\usepackage{natbib}
\usepackage{url}
\usepackage{multirow}
\usepackage{caption}
\usepackage{hyperref}
\usepackage{color}
\usepackage{booktabs}
\usepackage[margin=1in]{geometry}
\usepackage{authblk}
\usepackage{hhline}
\usepackage{algorithmic,algorithm}
\usepackage{scalerel}
\usepackage{soul}

\newtheorem{theorem}{Theorem}

\newtheorem{proposition}{Proposition}
\newtheorem{remark}{Remark}
\newtheorem{corollary}{Corollary}
\newtheorem{definition}{Definition}
\newtheorem{assumption}{Assumption}

\def\trans{^{\scriptscriptstyle \sf T}}

\def\pr{\mathsf{P}} 

\begin{document}

\title{A Decorrelating and Debiasing Approach to Simultaneous Inference for High-Dimensional Confounded Models}

\author{Yinrui Sun, Li Ma, and Yin Xia \\
    Department of Statistics and Data Science, Fudan University}

\date{}

\maketitle

\bigskip

\begin{abstract}
	Motivated by the simultaneous association analysis with the presence of latent confounders, this paper studies the large-scale hypothesis testing problem for the high-dimensional confounded linear models with both non-asymptotic and asymptotic false discovery control. 
	Such model covers a wide range of practical settings where both the response and the predictors may be confounded.
	In the presence of the high-dimensional predictors and the unobservable confounders, the simultaneous inference with provable guarantees becomes highly challenging, and the unknown strong dependence among the confounded covariates makes the challenge even more pronounced.
	This paper first introduces a decorrelating procedure that shrinks the confounding effect and weakens the correlations among the predictors, then performs debiasing under the decorrelated design based on some biased initial estimator.
	Following that, an asymptotic normality result for the debiased estimator is established and standardized test statistics are then constructed.
	Furthermore, a simultaneous inference procedure is proposed to identify significant associations, and both the finite-sample and asymptotic false discovery bounds are provided. 
	The non-asymptotic result is general and model-free, and is of independent interest.
	We also prove that, under minimal signal strength condition, all associations can be successfully detected with probability tending to one.
	Simulation and real data studies are carried out to evaluate the performance of the proposed approach and compare it with other competing methods.

\end{abstract}


\newpage

\baselineskip=20pt

\section{Introduction}\label{intro.sec}

The hidden confounding model has been widely adopted in many fields such as biology, medical science and economics. In the meanwhile, statistical inference that incorporates unobservable confounders arises frequently from a wide variety of applications including gene expression analysis, epidemiology studies and firm revenue predictions \citep[e.g.,][]{hsu2012reducing, sheppard2012confounding, sila2016women}. 
For example, to identify differentially expressed genes with respect to different disease states, confounding may occur in the form of batch effects and the confounders may include processing dates or other unrecorded laboratory conditions \citep{leek2010tackling}.
On the other hand, the confounded model may appear in the forms of perturbed model, measurement error model and linear structural equation model as reported and discussed in \cite{guo2022}. 
For instance, to quantify the impact of particulate matters and gaseous pollutants on human mortality rate, measurement errors of some pollutants may serve as confounding effects \citep{schwartz2003control}.

Under the confounded framework, this article addresses the problem of simultaneous association analysis between a possibly confounded response of interest and a set of high-dimensional predictors that may be affected by the confounding factors as well.
Specifically, under the following confounded linear model
\begin{equation}\label{model_confounder1}
	Y= X\beta + H\phi+ \xi, 
\end{equation}
where $Y \in \mathbb{R}^n$, $X \in \mathbb{R}^{n\times p}$, $H \in \mathbb{R}^{n\times q}$ and $\xi \in\mathbb{R}^n$ respectively represent the random response, predictors, latent confounders and the noise, while 
$\beta\in\mathbb{R}^p$ and $\phi\in\mathbb{R}^q$ respectively represent the deterministic coefficients of the predictors and the confounders,
we aim at simultaneous testing of hypotheses
\begin{equation}\label{test}
\mathcal{H}_{0,j}:\beta_j = 0 \text{ versus } \mathcal{H}_{1,j}:\beta_j \neq 0 ,~j=1,\ldots,p,
\end{equation}
with proper error rate control. 
In the presence of the unobservable $H$ and the high-dimensional $X$, provable statistical analysis for \eqref{test} becomes highly challenging. 

Based on Model \eqref{model_confounder1}, we further adopt the factor model structure to characterize the dependence between predictors and confounders \citep{cevid2020,guo2022}, i.e.,
\begin{equation}\label{model_factor}
	X= H\Psi+E,
\end{equation}
where $\Psi\in\mathbb{R}^{q\times p}$ and $E \in \mathbb{R}^{n\times p}$ respectively represent the deterministic loading matrix and the idiosyncratic error.
This model may lead to highly correlated predictors when a large number of covariates are affected by the latent confounders.
Such a property makes it a valuable tool, for example, for modeling highly correlated genes in the presence of unmeasured confounders in genetic analysis \citep{leek2007capturing,leek2008general}.
However, the strong dependence poses technical difficulties for false discovery control, making the challenge of the multiple testing problem \eqref{test} even more pronounced.

\subsection{Connections to Existing Works}\label{sec:literature}

For linear models without confounders, Lasso \citep{tibshirani1996} related methods are popularly employed for consistent estimation and prediction \citep[e.g.,][]{bickel2009, negahban2012} as well as variable selection \citep[e.g.,][]{zhao2006,lahiri2021}.
However,  Lasso often produces biased estimators \citep{fan2001,zou2006adaptive}, and debiased procedures are developed for the inference of a single coefficient \citep[e.g.,][]{zhang2014,javanmard2014,van2014}.
Such asymptotically unbiased estimators can be further adopted to test the coefficient vector globally with maximum-type statistics \citep{zhang2017,dezeure2017}, or to test local components simultaneously with FDR control \citep{javanmard2019}.
In addition, an alternative inverse regression approach is proposed by \citet{liu2014} for multiple testing of linear regression coefficients and it is further extended to the comparison of two high-dimensional linear regression models by \cite{xia2018}. 
Nevertheless, the aforementioned approaches are restricted to the conventional linear regression models and cannot be easily extended to the confounded settings.

More recently, linear models that incorporate hidden confounders have been explored. 
\citet{chernozhukov2017} combines Lasso and Ridge penalties to estimate a perturbed model that contains the confounded linear model as a special setting.
To reduce the confounding effect, \citet{cevid2020} discusses a class of spectral transformations and proposes a Trim transformation to estimate the sparse vector $\beta$ in \eqref{model_confounder1}.
Under the factor model structure \eqref{model_factor}, \citet{guo2022} further employs the Trim transform and performs inference on a single component of $\beta$ through Lasso debiasing.
Moreover, multivariate linear regression with hidden confounders is studied by \citet{bing2022estimation,bing2022inference}.
Specifically, \citet{bing2022estimation} aims at estimation and proposes a subspace projection method to reduce the confounding effect, while \citet{bing2022inference} performs additional statistical inference on a single {entry} of the coefficient matrix.
To the best of our knowledge, no existing literatures have addressed the large-scale inference problem \eqref{test}, and the existing technical tools that aim at  single coefficient inference cannot be directly applied to the simultaneous inference problem with strong dependence as shown in the current confounded framework.

\subsection{Challenges of Dependence}\label{sec:dependency}

The false discovery proportion (FDP) and false discovery rate (FDR) \citep{benjamini1995controlling} provide a powerful and practical criterion for test error evaluation in large-scale inference. 
The proposed Benjamini-Hochberg (BH) procedure controls FDR under positive regression dependence and achieves asymptotic error control for weakly dependent $p$-values \citep{benjamini2001control,storey2004strong}.
To deal with arbitrary dependence, several approaches are developed \citep[e.g.,][]{benjamini2001control,wang2022false}, which may however suffer from some power loss \citep{fithian2022conditional}.
In another line of research, \citet{efron2007correlation} emphasizes the issue of consistent FDR estimation under dependence; \cite{fan2012estimating} and \cite{fan2017estimation} develop methods to estimate FDP consistently by exploiting the covariance structures of dependent Gaussian test statistics.
In light of these achievements, developing provably valid and powerful methods for simultaneous testing in complex statistical models with strong dependence remains a challenging task.

\subsection{Our Contributions}
In this article, we fill the methodological and theoretical gap of large-scale inference on high-dimensional linear regression models that can accommodate both the presence of latent confounders and highly correlated predictors.
A key decorrelating transformation is introduced and it simultaneously diminishes the confounding effect and reduces the strong dependence among the covariates. 
Following that, a component-wise debiasing procedure based on the decorrelated design and some initial estimators is proposed. 
After decorrelating and debiasing, the confounding effect is shown to be asymptotically negligible and the dependence among the debiased estimators is derived. 
Furthermore, the proposed debiased estimator is shown to be square-root $n$ consistent and asymptotic normal.
Finally, the test statistics for all $\beta_j$'s are constructed and a multiple testing procedure is proposed to simultaneously detect significant associations with both asymptotic and non-asymptotic false discovery rate control.
The proposed procedure is shown to enjoy full power asymptotically under some minimal signal strength condition.

Our proposal makes novel and useful contributions from both the methodological and theoretical perspectives. 
Methodologically, the decorrelating transformation is novelly applied as a preprocessing step of the subsequent debiasing procedure.
Note that, such transformation is fundamentally different from the factor-adjusted procedure of \citet{fankewang2020} though they both reduce high correlations. 
\citet{fankewang2020} aims at model selection consistency of generalized linear models with factor model structure, and decorrelates the original linear models so that the new design matrix satisfies the irrepresentable condition. 
In comparison, we aim at simultaneous testing of confounded linear models and the decorrelating function is applied in the debiasing step so that the confounding effects can be weakened and the dependence among test statistics can be characterized. 

Second, novel test statistics are built upon new debiasing procedures.
Such construction is closely related to the Trim-type test statistics \citep{guo2022}, but is far from trivial extensions.
To perform statistical inference on the single component of $\beta$, asymptotic normality property of their statistics is established in \citet{guo2022}, while the dependence among their test statistics remains unknown. 
Hence, it is unclear whether the multiple testing procedure based on their statistics is theoretically guaranteed. 
In contrast, by the new debiasing strategy,
the dependence structure of the proposed test statistics is explicitly calculated and the false discovery bounds of the corresponding test are explored.
In addition, compared to the test based on \citet{guo2022}, the proposed method enjoys computational advantage and improved signal-noise ratio; see Remark \ref{remark:gamma_j} and Section \ref{sec:compare}.
The numerical studies affirm the superior performance of our method.

Theoretically, first of all, a brand new finite sample analysis of FDR control is established. 
Such non-asymptotic result is model-free and it can be applied to various simultaneous inference problems on testing of high-dimensional covariance matrices, graphical models, etc, and is of independent interest. 
It is among the first results that study the non-asymptotic FDR control for the multiple testing procedure performed in this article. 
Second, a modified Cramer-type Gaussian approximation result that improves the one in \citet{liu2013,cai2016} is presented and proved in Lemma 7 of the supplement, and it makes a useful addition to the general toolbox of asymptotic simultaneous inference.

\subsection{Outline of the Paper}

The rest of the paper is organized as follows. 
Section \ref{sec:algo} introduces the model and outlines the proposed method.
Section \ref{sec:debiased_est} studies the construction of the test statistics. It starts with the decorrelating transformation and then performs debiasing based on the decorrelated design. Theoretical properties are collected in Section \ref{sec:theory-stat}, {and methodology comparison is discussed in Section \ref{sec:compare}}.
Section \ref{sec:multi_testing} proposes a large-scale inference procedure and explores its finite-sample and asymptotic FDR control results and power analysis.  
Simulations and real data analysis are provided in Sections \ref{sec:simu} and \ref{sec:real_data}.
The technical proofs, assumption verifications {and additional numerical explorations} are collected in the supplementary material.

\section{Model and Algorithm}\label{sec:algo}
This section outlines the key steps of the proposed method. We start with some notation that will be used throughout the paper and then introduce the confounded linear model.

Denote by $[n] = \left\{1,2,\ldots,n\right\}$ for a positive integer $n$. 
Denote the cardinality of a set $S$ by $|S|$.
For $x\in\mathbb{R}$, denote by $\lfloor x \rfloor$ the largest integer  no larger than $x$. 
Let $a \vee b = \max(a,b)$ and $a \wedge b = \min(a,b)$. 
For $a\in\mathbb{R}^n$, let $\text{supp}\left(a\right) = \left\{j\in[n]:a_j\neq0\right\}$;
let $a_{\scriptscriptstyle S} = (a_i)_{\scriptscriptstyle i\in S} \in\mathbb{R}^{\scriptscriptstyle |S|}$, and $a_{\scriptscriptstyle -S} = a_{\scriptscriptstyle  [n]\backslash S }\in\mathbb{R}^{\scriptscriptstyle n-|S|}$ for any index set $S \subset [n]$. For $A\in\mathbb{R}^{m\times n}$, denote by $A_{i,\cdot}$ and $A_{\cdot,j}$ the $i$-th row and $j$-th column of $A$, respectively, and denote by $A_{\cdot,-j}\in\mathbb{R}^{m\times(n-1)}$ the submatrix of $A$ with the $j$-th column removed. Let $\Lambda_i\left(A\right)$ be the $i$-th largest singular value of $A$, and let $\left|A\right|_\infty = \max_{i\in[m],j\in[n]}\left|A_{i,j}\right|$.
For two positive sequences $\{a_n\}$ and $\{b_n\}$, write $a_n \lesssim b_n$ if there exists some constant $C>0$ such that $a_n\leq Cb_n$ for all $n$, and $a_{n} \asymp b_{n}$ if $a_n \lesssim b_n$ and $b_n \lesssim a_n$.
{For two  positive random sequences $\{X_n\}$ and $\{Y_n\}$, write $X_n \lesssim_\pr Y_n$ if $X_n \lesssim Y_n$ holds with probability tending to 1 as $n\rightarrow\infty$.}
Define $\|X\|_{\psi_p} = \inf\left\{ t>0: \mathbb{E}\psi_p\left( |X|/t\right) \leq 1 \right\}$ with $\psi_p(x) = \exp\left(x^p\right) - 1,p\geq 1$.
A random variable $X$ is sub-Gaussian if $\|X\|_{\psi_2}<\infty$, and is sub-exponential if $\|X\|_{\psi_1}<\infty$;
$\|X\|_{\psi_2}$ and $\|X\|_{\psi_1}$ are called sub-Gaussian norm and sub-exponential norm, respectively.
A random vector $X\in\mathbb{R}^n$ is sub-Gaussian if $x\trans X$ is sub-Gaussian for any $x\in\mathbb{R}^n$, and its sub-Gaussian norm is defined as $\|X\|_{\psi_2} = \sup_{\|x\|_2 = 1}\|x\trans X\|_{\psi_2}$. 
Denote by $C,c,C_1,c_1, \ldots$ some universal positive constants that are independent of the sample size $n$, and may differ from place to place.

\subsection{Model Description}\label{sec:model}
Recall that, provided the independent and identically distributed (i.i.d.) observed data $\left\{X_{i, \cdot}, Y_{i}\right\}_{ i \in [n]}$ and the unobserved latent confounders $\left\{H_{i, \cdot}\right\}_{i \in [n]}$, the confounded linear model \eqref{model_confounder1} with the factor model structure \eqref{model_factor} is given by
\begin{equation}\label{model_confounder2}
	Y_{i}= X_{i, \cdot}\beta + H_{i, \cdot}\phi +\xi_i \text { and } X_{i, \cdot} = H_{i, \cdot}\Psi + E_{i, \cdot}, \text{ for }i\in[n].
\end{equation}
Throughout it is assumed that the regression vector $\beta$ is sparse and the number of confounders is smaller than the number of predictors and the sample size, i.e., $q< (n\wedge p)$.
We further assume that the idiosyncratic error $E\trans_{i,\cdot}\in \mathbb{R}^p$ has zero mean and covariance $\Sigma_E>0$, and is independent with the latent factor $H\trans_{i, \cdot}\in \mathbb{R}^q$; the random error $\xi_i \in \mathbb{R}$ has zero mean and variance $\sigma_\xi^2>0$, and is independent with $\left(H_{i, \cdot}, E_{i, \cdot}\right){\trans} \in \mathbb{R}^{q+p}$.
By the fact that $H_{i, \cdot}\Psi = H_{i, \cdot}O\trans O\Psi$ holds for any $q\times q$ orthonormal matrix $O$, 
{without loss of generality,} 
it is assumed that $\mathbb{E}\left(H_{i,\cdot}\right) =0$, Cov$(H\trans_{i,\cdot}) := \Sigma_H = I_q$ and $\Psi\Psi\trans$ is diagonal.
Hence the covariance matrix of $X\trans_{i,\cdot}$ is given by
\begin{equation}\label{covariance_x}
	\Sigma_X = \Psi\trans\Psi + \Sigma_E.
\end{equation}
Under the dense confounding scenarios where a large number of predictors are affected by the latent factors, the covariance $\Sigma_X$ is spiked and predictors $X_{i,\cdot}$ are strongly correlated with each other. Such phenomena may lead to a highly challenging simultaneous inference on $\mathcal{H}_{0,j},j\in[p]$.

Moreover, the confounding effect on $Y_{i}$ introduced by $H_{i, \cdot}$ adds difficulties on the large-scale inference. 
To better interpret such effect, we project $H_{i,\cdot} \phi$ onto $X_{i,\cdot}$ linearly in the $L_2$ sense:
$
H_{i,\cdot}\phi = X_{i,\cdot}b + \left(H_{i,\cdot}\phi - X_{i,\cdot}b\right),
$
in which the non-correlation $\text{Cov}\left(X\trans_{i,\cdot}, H_{i,\cdot}\phi - X_{i,\cdot}b\right) = 0$ yields that
$b = \Sigma_X^{-1}\text{Cov}(X\trans_{i,\cdot},H\trans_{i,\cdot})\phi$. 
Hence the confounded model \eqref{model_confounder2} can be represented as a linear regression model with a perturbed coefficient $b$, i.e.,
\begin{equation}\label{model_perturb2}
Y = X\left(\beta + b\right)+ \varepsilon \text{ and } X = H\Psi + E,
\end{equation}
where $Y = (Y_1,\ldots,Y_n)\trans$, $X = (X_{1,\cdot}\trans,\ldots,X_{n,\cdot}\trans)\trans$, $H = (H_{1,\cdot}\trans,\ldots,H_{n,\cdot}\trans)\trans$ , $E = (E_{1,\cdot}\trans,\ldots,E_{n,\cdot}\trans)\trans$, 
$
b = \Sigma_X^{-1}\Psi\trans\phi = \left(\Psi\trans\Psi + \Sigma_E\right)^{-1}\Psi\trans\phi,
$
and the error term $\varepsilon = (\varepsilon_1,\cdots,\varepsilon_n)\trans = \xi + \Delta$, in which $\Delta = (\Delta_1,\cdots,\Delta_n)\trans$ with
$
\Delta_i := H_{i,\cdot}\phi - X_{i,\cdot}b = H_{i,\cdot}\left(\phi - \Psi b\right) - E_{i,\cdot}b.
$
Elemental calculation yields that the variance of $\varepsilon_i$ equals to 
$
\sigma_\varepsilon^2 := \sigma_\xi^2+\text{Var}\left(\Delta_i\right) = \sigma_\xi^2 + \phi\trans\left(I_q - \Psi\Sigma_X^{-1}\Psi\trans\right)\phi.
$
The perturbation term $Xb$ in \eqref{model_perturb2} can be viewed as the confounding effect in the presence of $H$.
The bias induced by such perturbation may lead to the detection of spurious associations and therefore poses additional difficulties on the simultaneous inference.

\subsection{Outline of the Algorithm}\label{sec:outline}

To overcome the aforementioned difficulties, we propose in this article a decorrelating and debiasing approach to the simultaneous inference \eqref{test} on high-dimensional confounded linear models.
We first outline the steps of the proposed procedure in Algorithm \ref{alg1} and then study the details of each key step later in Sections \ref{sec:decorrelate} to \ref{sec:test-stat} and Section \ref{sec:multiple-test}.

\begin{algorithm}[h]

\caption{\label{alg1} A Decorrelating and Debiasing Approach to Simultaneous Inference.}
{Input: }$X \in \mathbb{R}^{n \times p}$, $Y \in \mathbb{R}^n$, initial estimators $\{\widehat{\beta}_j, j\in [p]\}$ and $\widehat\sigma_\xi^2$, and tuning parameters $\{\lambda_j, j\in [p]\}$.
\begin{enumerate}
\item[1.] \label{alg1:step1} {\bf Decorrelating}: 
obtain decorrelating transformed $F_{\text{dc}} X$ and $F_{\text{dc}} Y$, where
\[
	F_{\text{dc}} = \sum_{i=1}^n d_i u_i u_i\trans \text{ with } d_i = \mathbb{I}\left(i>q\right), \text{ and $u_i$'s are left singular vectors of $X$.} 
\]

\item[2.] \label{alg1:step2} {\bf Debiasing}:
for $j\in [p]$, debias the initial biased estimator $\widehat{\beta}_j$ by
\[
	\overline{\beta}_j =  \widehat{\beta}_j + \frac{z_j\trans\left(F_{\text{dc}}Y - F_{\text{dc}}X\widehat{\beta}\right)}{z_j\trans F_{\text{dc}}X_{\cdot,j}}, 
\]
where $	z_j = F_{\text{dc}}X_{\cdot,j} - F_{\text{dc}}X_{\cdot,-j}\widehat{\gamma}_j$, $D = \text{diag}\left( \|F_{\text{dc}}X_{\cdot,1}\|_2,\ldots,\|F_{\text{dc}}X_{\cdot,p}\|_2 \right) / \sqrt{n}$ and $\widehat{\gamma}_j = \arg\min_{\gamma\in\mathbb{R}^{p-1}}\left\{\frac{1}{2n}\|F_{\text{dc}}X_{\cdot,j} - F_{\text{dc}}X_{\cdot,-j}\gamma\|_2^2 + \lambda_{j}\|D_{-j,-j}\gamma\|_1\right\}$.

\item[3.] \label{alg1:step3} {\bf Obtain test statistics}: 
for $j\in [p]$, construct the test statistic
\[
	{T_j = \frac{\sqrt{n}\overline{\beta}_j}{\widehat\sigma_\xi \tau_j}, \text{ with } \tau_j = \sqrt{n}\| z_j \|_2^{-1}.
	}
\]

\item[4.] \label{alg1:step4} {\bf Simultaneous inference}: perform multiple testing by Algorithm \ref{alg2} in Section \ref{sec:multiple-test} and obtain the rejection set.
\end{enumerate}
\end{algorithm}

\section{Construction of the Test Statistics}\label{sec:debiased_est}
This section first studies the detailed steps for constructing the test statistics in Algorithm \ref{alg1}. 
Then the theoretical properties of the decorrelated design as well as the debiased estimators are presented in Section \ref{sec:theory-stat}.
{In Section \ref{sec:compare}, we compare the proposed method with \citet{guo2022}.}

\subsection{Step 1: Decorrelating}\label{sec:decorrelate}
Based on the model described in Section \ref{sec:model}, we introduce the first step of our statistic construction: decorrelation. 
We start with reviewing the class of spectral transformation proposed in \citet{cevid2020}.
Let $X = \sum_{i=1}^{n\wedge p}\Lambda_i\left(X\right)u_iv_i\trans$ be the singular value decomposition (SVD) of the design matrix $X$, where $\Lambda_1\left(X\right)\geq \Lambda_2\left(X\right)\geq\cdots\geq \Lambda_{n\wedge p}\left(X\right)\geq0$.
For the dense confounding scenarios, the spiked covariance \eqref{covariance_x} would produce top $q$ singular values that are relatively large, which makes the perturbation term $Xb$ in \eqref{model_perturb2} not negligible in $\ell_2$ norm sense.
To reduce the confounding effect, \citet{cevid2020} considers a class of spectral transform matrix
$
F = \sum_{i=1}^n d_i u_i u_i\trans,
$
and $X$ is left-multiplied by $F$ to reach a preconditioned design $FX = \sum_{i=1}^{n\wedge p}d_i\Lambda_{i}(X)u_iv_i\trans$. 
Proper choice of $d_i$ shrinks top singular values of $X$ to a moderate level and eliminates the spiked structure. 
In particular, \citet{cevid2020} introduces the Trim transform:
\begin{equation}\label{F_trim}
F_{\text{T}} = \sum_{i=1}^n d_i u_iu_i\trans, \text{ with } d_i = \frac{\Lambda_{\lfloor \rho (n\wedge p) \rfloor }\left(X\right)}{\Lambda_i\left(X\right)}\mathbb{I}\left(i\leq\lfloor \rho (n\wedge p) \rfloor\right)+\mathbb{I}\left(i>\lfloor \rho (n\wedge p) \rfloor\right),
\end{equation}
where $\rho\in\left(0,1\right)$ is some preassigned parameter satisfying $\lfloor \rho (n\wedge p) \rfloor \geq q+1$. Then $F_{\text{T}}Y$ is regressed on $F_{\text{T}}X$ through Lasso procedure to estimate the sparse coefficient $\beta$.
Next, to reduce bias and perform inference on a single component $\beta_j$, \citet{guo2022} applies the Trim transform as well as the debiased Lasso method in \citet{zhang2014} to construct an asymptotically unbiased estimator of $\beta_j$. Nevertheless, the dependence structure among the covariates after such transformation is unknown, and hence the multiple testing problem \eqref{test} remains a challenging task.

To achieve the goal of reducing the bias introduced by the confounding effects as well as weakening the dependence among the predictors for the subsequent simultaneous inference, motivated by \citet{fan2013,fan2018},
we propose the following decorrelating function,
\begin{equation}\label{F_decorrelation}
F_{\text{dc}} = \sum_{i=1}^n d_i u_i u_i\trans \text{ with } d_i = \mathbb{I}\left(i>q\right).
\end{equation}
Such transformation provides a projection matrix onto the subspace spanned by $\left(u_{q+1},\cdots,u_{n}\right)$. 
For simplicity, we assume $q$ is known in Algorithm \ref{alg1:step1} and the following theoretical analysis.
Nevertheless, the number of hidden confounders is unknown in practice, and rich methods have been developed to determine $q$ consistently \citep[e.g.,][and references therein]{bai2002,ahn2013,fan2022}.
In fact, the theories in Sections \ref{sec:theory-stat} and \ref{sec:multi_testing} remain {valid} if $q$ is replaced by an estimate $\widehat{q}$ that satisfies $\mathbb{P}\left( \widehat{q} = q \right) \rightarrow 1$.
We also investigate the numerical impact of misspecified $q$ in Section E.4 of the supplement. 
The results indicate that an overestimation of $q$ has little impact on the performance of the proposed method, which agrees with the findings in \citet{fan2013}. Hence a slightly larger $\widehat q$ is preferred in practice.

To interpret $F_{\text{dc}}$, it will be shown in Proposition \ref{factor_estimation_bound} of Section \ref{sec:theory-stat} that, the decorrelating transformed design matrix, i.e., $F_{\text{dc}}X = \sum_{i=q+1}^{n\wedge p}\Lambda_i\left(X\right)u_iv_i\trans$, is close to the idiosyncratic error $E$.
Hence, the top singular values of $F_{\text{dc}}X$ are compressed compared to those of $X$ and the dependence structure of $F_{\text{dc}}X$ can be approximated by the counterpart of $E$.
Therefore, $F_{\text{dc}}$ serves as a shrinkage method to reduce the confounding effect $Xb$ as well as a decorrelation approach to yield weakly correlated design, and is thus important in the debiasing step as well as the following multiple testing procedure.

\begin{remark}\label{remark_decorrelate}
Before the rigorous analysis in Proposition \ref{factor_estimation_bound}, we provide one insight that $F_{\text{dc}}X$ is close to $E$.
To estimate the latent confounders $H$ and the loading matrix $\Psi$, one may consider the following constrained optimization problem \citep{stock2002,bai2002}:
\begin{equation}\label{factor_estimation}
	\min_{H\in\mathbb{R}^{n\times q},~\Psi\in\mathbb{R}^{q\times p}}\|X - H\Psi\|_F^2, \text{ s.t. } \frac{1}{n}H\trans H = I_q,
\end{equation}
and the solution is given by
$
\widehat{H} = \sqrt{n}\left(u_1,\cdots,u_q\right),~\widehat{\Psi}\trans=\frac{1}{n}X\trans \widehat{H} = \frac{1}{\sqrt{n}}\left\{\Lambda_1(X)v_1,\cdots,\Lambda_q(X)v_q\right\}.
$
Therefore we have $F_{\text{dc}}X = \sum_{i=q+1}^{n\wedge p}\Lambda_i\left(X\right)u_iv_i\trans = X - \widehat{H}\widehat{\Psi}$, which approximates $E$.
\end{remark}

\subsection{Step 2: Debiasing}\label{sec:debias}

Based on the decorrelating function $F_{\text{dc}}$ defined in \eqref{F_decorrelation}, analogously to that in \citet{zhang2014}, the debiased estimator $\overline{\beta}_j$ under the decorrelated design can be obtained by
\begin{equation}\label{beta_debias}
\overline{\beta}_j = \widehat{\beta}_j + \frac{z_j\trans\left(F_{\text{dc}}Y - F_{\text{dc}}X\widehat{\beta}\right)}{z_j\trans F_{\text{dc}}X_{\cdot,j}} 
	=  \frac{z_j\trans\left(F_{\text{dc}}Y - F_{\text{dc}}X_{\cdot,-j}\widehat{\beta}_{-j}\right)}{z_j\trans F_{\text{dc}}X_{\cdot,j}},
\end{equation}
where $\widehat{\beta}$ is some initial biased estimator and $z_j$ is a suitably chosen vector so that the bias of $\widehat{\beta} - \beta$ can be reduced to an acceptable level for the subsequent analysis. 

It will be shown in Section \ref{sec:theory-stat} that any initial estimator $\widehat{\beta}$ that satisfies Assumption \ref{con_initial} can reach the desired theoretical properties and hence can be employed in the construction of $\overline{\beta}_j$.
As an example, we can apply some spectral transform $F$ together with Lasso procedure to obtain a biased initial estimator $\widehat{\beta}$ \citep{cevid2020,guo2022}.
Specifically, one regresses $FY$ on $FX$ and obtains $\widehat{\beta}$ as follows:
\begin{equation}\label{beta_lasso}
	\widehat{\beta} = \arg\min_{\beta\in\mathbb{R}^p}\left\{\frac{1}{2n}\|FY - FX\beta\|_2^2 + \lambda\|\widetilde{D}\beta\|_1\right\},
\end{equation}
where $\widetilde{D} = \text{diag}\left( \|FX_{\cdot,1}\|_2,\ldots,\|FX_{\cdot,p}\|_2 \right) / \sqrt{n} \in \mathbb{R}^{p\times p}$, and the function $F$ can be either the Trim transform in \eqref{F_trim} or the decorrelation transform in \eqref{F_decorrelation}. 
It is shown in Proposition 4 in Section D.2 of the supplement that such estimators indeed satisfy Assumption 6.

Next, to obtain a desired $z_j$, node-wise Lasso \citep{meinshausen2006} that regresses $F_{\text{dc}}X_{\cdot,j}$ on $F_{\text{dc}}X_{\cdot,-j}$ is performed first and the corresponding regression coefficient estimate $\widehat{\gamma}_j$ is calculated by
\begin{equation}\label{gamma_j_lasso}
	\widehat{\gamma}_j = \arg\min_{\gamma\in\mathbb{R}^{p-1}}\left\{\frac{1}{2n}\|F_{\text{dc}}X_{\cdot,j} - F_{\text{dc}}X_{\cdot,-j}\gamma\|_2^2 + \lambda_{j}\|D_{-j,-j}\gamma\|_1\right\},
\end{equation}
where $D = \text{diag}\left( \|F_{\text{dc}}X_{\cdot,1}\|_2,\ldots,\|F_{\text{dc}}X_{\cdot,p}\|_2 \right) / \sqrt{n} \in \mathbb{R}^{p\times p}$.
Because $F_{\text{dc}}X$ is well approximated by $E$ as claimed in Proposition \ref{factor_estimation_bound},
such estimate $\widehat{\gamma}_j$ resembles the true regression vector $\gamma_j$ that regresses $E_{i,j}$ on $E_{i,-j}$, i.e., 
\begin{equation}\label{gamma_j}
		\gamma_j =  \arg\min_{\gamma\in\mathbb{R}^{p-1}}\mathbb{E}\left(E_{i,j} - E_{i,-j}\gamma \right)^2
		= \left(\Sigma_{E,-j,-j}\right)^{-1}\Sigma_{E,-j,j}
		=  -\Omega_{E,-j,j}/\Omega_{E,j,j},
\end{equation}
where $(\omega_{k,j}) =: \Omega_E = \Sigma_E^{-1}$ is the precision matrix of the error $E$.
It will be rigorously shown in Proposition \ref{gammaj_error} of Section \ref{sec:theory-stat} that, $\gamma_j$ is indeed well approximated by the Lasso estimator $\widehat{\gamma}_j$.
Finally, the vector $z_j\in\mathbb{R}^n$ is constructed as
\begin{equation}\label{z_j}
	z_j = F_{\text{dc}}X_{\cdot,j} - F_{\text{dc}}X_{\cdot,-j}\widehat{\gamma}_j.
\end{equation}
It follows that $z_j$ well resembles the true residual $\eta_j$ as defined below:
\begin{equation}\label{eta_ij}
	\eta_j = \left(\eta_{1,j},\cdots,\eta_{n,j}\right)\trans, \text{ with }\eta_{i,j} = E_{i,j} - E_{i,-j}\gamma_j,\quad i\in [n].
\end{equation}

\subsection{Step 3: Constructing Test Statistics}\label{sec:test-stat}

After decorrelating and debiasing, we next construct the test statistics for $\mathcal{H}_{0,j},j\in[p]$.
As will be shown in the proof of Theorem \ref{asymp_normal}, $\sqrt{n}\left( \overline{\beta}_j - \beta_j \right) = \frac{\omega_{j,j}}{\sqrt{n}}\eta_j\trans \xi + \text{Re}_j$, where $\text{Re}_j$ is an asymptotically negligible remainder, and the scaled sum of the influence functions $\frac{\omega_{j,j}}{\sqrt{n}}\eta_j\trans \xi$ converges weakly to $N\left( 0, \sigma_\xi^2 \omega_{j,j}\right)$.
To approach the unknown $\omega_{j,j}$, by the fact that $\mathbb{E}\eta_{i,j}^2 = 1/\omega_{j,j}$, one can estimate $\sqrt{\omega_{j,j}}$ by $\sqrt{n} / \| z_j \|_2$ because $z_j$ well resembles $\eta_j$.

Therefore, the final standardized test statistic for $\mathcal{H}_{0,j}:\beta_j = 0$ that accommodates the component-wise heteroscedasticity is obtained by
\begin{equation}\label{T_j}
	T_j = \frac{\sqrt{n}\overline{\beta}_j}{\widehat\sigma_\xi \tau_j} \text{ with } \tau_j = \frac{\sqrt{n}}{\| z_j \|_2}, ~j \in [p],
\end{equation}
where $\widehat\sigma_\xi$ is some consistent estimator of the standard deviation $\sigma_\xi$ of the random error $\xi$. 
For the initial estimator in \eqref{beta_lasso}, the estimate of $\sigma_\xi^2$ can be obtained by
\begin{equation}\label{sigma_hat}
	\widehat\sigma_\xi^2 = \|FY - FX\widehat{\beta}\|_2^2 / \text{Tr}(F\trans F),
\end{equation}
where Tr$(\cdot)$ denotes the trace of a matrix.
Similarly as the discussion on the initial estimator $\widehat\beta$, any estimator $\widehat\sigma_\xi^2$ that satisfies Assumption \ref{con_initial} can reach the desired theoretical properties. 
Again, it is shown in Proposition 4 of the supplement that the estimator \eqref{sigma_hat} indeed satisfies this assumption.
However, the practical choice of $\lambda$ in \eqref{beta_lasso} is usually determined by the data-driven cross-validation strategy, which empirically produces overfitting coefficients that lead to an underestimation of the error variance.
Inspired by the suggestion in \citet{reid2016study} as well as its application in global testing problem \citep{zhang2017}, we propose a new calibrated noise variance estimator for practical purpose; see the detailed construction and explanations in Section E.2 of the supplement.

\subsection{Theoretical Properties}\label{sec:theory-stat}

This section studies the theoretical properties related to the aforementioned key steps for the test statistic construction. We first collect some definitions and technical assumptions. 
\begin{definition}[Restricted Eigenvalue, \cite{bickel2009}]\label{re_def}
	For some matrix $A\in\mathbb{R}^{n\times p}$ and scalars $s,L>0$, the restricted eigenvalue $\kappa\left(A, s, L\right)$ is defined as
	$$
	\kappa\left(A, s, L\right) = \min_{S\subset[p]\atop|S|\leq s}\min_{x\in\mathbb{R}^{\scaleto{p}{2.5pt}}:~x\neq0 \atop \|x_{\scaleto{S^c}{3pt}}\|_1\leq L\|x_{\scaleto{S}{2.5pt}}\|_1}\frac{\|Ax\|_2}{\sqrt{n}\|x_{\scriptscriptstyle S}\|_2}.
	$$
\end{definition}

\begin{assumption}\label{con_E_H}
	\noindent
	\begin{enumerate}
		\item $\Sigma_E^{-1/2}E_{i,\cdot}\trans$ is sub-Gaussian with bounded sub-Gaussian norm $v_E:=\|\Sigma_E^{-1/2}E_{i,\cdot}\trans\|_{\psi_2}$, and there exists a constant $C_E>0$ such that $C_E^{-1}\leq \Lambda_p\left(\Sigma_E\right)\leq\Lambda_1\left(\Sigma_E\right)\leq C_E$.
		
		\item $H\trans_{i,\cdot}$ is sub-Gaussian with bounded sub-Gaussian norm $v_H:=\|H\trans_{i,\cdot}\|_{\psi_2}$.
	\end{enumerate}
\end{assumption}

\begin{assumption}\label{con_Psi}
	There exists a  constant $C_{\Psi}>0$ such that $|\Psi|_\infty\leq C_\Psi$, and $C_\Psi^{-1} \leq \frac{1}{\sqrt{p}}\Lambda_{q}(\Psi) \leq \frac{1}{\sqrt{p}}\Lambda_{1}(\Psi) \leq C_\Psi$.
\end{assumption}

\begin{assumption}\label{con_re_decorrelation}
Let $s_j = \big|\left\{k\in[p]:~\omega_{k,j}\neq 0\right\}\big|$ and $s_\Omega = \max_{j\in[p]}s_j$.
{For some sufficiently large constant $C>0$}, there exists a constant $\kappa_0>0$ such that
$
\kappa\left(F_\text{dc}X, s_\Omega, CA_n\right)\geq\kappa_0
$
holds with probability tending to 1, where $A_n = |F_{\text{dc}}X - E |_\infty \vee 1$.
\end{assumption}

\begin{assumption}\label{con_E_independence}
	$\eta_{i,j}$ is independent of $E_{i,-j}$.
\end{assumption}

\begin{assumption}\label{con_e_phi}
	$\xi_i$ is sub-Gaussian with bounded sub-Gaussian norm $v_\xi:=\|\xi_i\|_{\psi_2}$, and {$\|\phi\|_2^2 \lesssim \log(p) $}.
\end{assumption}

\begin{assumption}\label{con_initial}
	Let  $s_0 = |S_0|$, $S_0=\text{supp}\left(\beta\right)$. The initial estimators $\widehat{\beta}$ and $\widehat\sigma_\xi$ satisfy that
	$$
	\|\widehat{\beta} - \beta\|_1  = O_\pr \left( qs_0 \sqrt{ \frac{\log(p)}{n} } \right) \text{ and } |\widehat\sigma_\xi - \sigma_\xi| = O_\pr \left( {\frac{1}{\sqrt{n}}} + qs_0\frac{\log(p)}{n} \right)  .
	$$
\end{assumption}

\begin{remark}
Assumption \ref{con_E_H} is mild: 
the sub-Gaussian conditions rule out heavy-tailed latent confounders $H$ and idiosyncratic errors $E$, and the eigenvalue condition on $\Sigma_E$ ensures a well-conditioned covariance matrix.
Assumption \ref{con_Psi} is common in the factor model literatures \citep[e.g.,][]{fan2013,fan2018}, and it indicates dense confounding effects \citep{guo2022}. 
The restricted eigenvalue condition in Assumptions \ref{con_re_decorrelation} is widely adopted in sparse linear regression \citep[e.g.,][]{bickel2009,van2009,negahban2012}.
Assumption \ref{con_E_independence} is a technical condition that is also required in \citet{guo2022}; it holds if $E_{i,\cdot}$ is Gaussian.
Assumption \ref{con_e_phi} relaxes the Gaussian noise condition in \citet{guo2022} and assumes that the $l_2$ norm of the confounding coefficient $\phi$ is not growing too fast.
Assumption \ref{con_initial} provides the required convergence rates for the initial estimators. 
The verifications of Assumptions \ref{con_re_decorrelation} and \ref{con_initial} are collected in Section D of the supplement.
\end{remark}

We start with the following proposition that shows how well $F_{\text{dc}}X$ approximates the error $E$ in the entry-wise deviation sense, so to verify the claim in Section \ref{sec:decorrelate} that the decorrelating procedure reduces the confounding effect and yields weakly correlated design.

\begin{proposition}\label{factor_estimation_bound}
	Under Assumptions \ref{con_E_H}-\ref{con_Psi}, if $n \lesssim p$, $q\log(n) = o(n)$ and  and $n\geq C\log(p)$ for some constant $C>0$, we have
	$$
	| F_{\text{dc}}X - E |_\infty \lesssim_\pr \frac{q^{3/2}\log(n)}{\sqrt{n}} + q\log(n)\sqrt{\frac{\log(p)}{n}}.
	$$
\end{proposition}

We remark that, to ensure a well approximation of $E$ by $F_{\text{dc}}X$, the low-rank component $H\Psi \in \mathbb{R}^{n\times p}$ needs a significant compression by the projection $F_{\text{dc}}$. Henceforth, we require the strength of the spiked signals to be comparable with the sample size $n$.
Such condition is mild in high dimensions, and is also imposed in \citet{guo2022}.

Next, based on the deviation result of $F_{\text{dc}}X$ and $E$, we show in the following proposition that, the true regression vector $\gamma_j$ {in \eqref{gamma_j}} that regresses $E_{i,j}$ on $E_{i,-j}$ is well approximated by the estimated regression vector $\widehat\gamma_j$ obtained via node-wise Lasso {\eqref{gamma_j_lasso}} that regresses $F_{\text{dc}}X_{\cdot,j}$ on $F_{\text{dc}}X_{\cdot,-j}$. 
Therefore, the $z_j$ constructed in \eqref{z_j} well resembles the true residual $\eta_j$ in \eqref{eta_ij} by combining Propositions \ref{factor_estimation_bound} and \ref{gammaj_error}.

\begin{proposition}\label{gammaj_error}
Under Assumptions \ref{con_E_H}-\ref{con_E_independence}, if $q=o(n)$, $n\geq C\log(p)$ and
$$
\lambda_{j} \geq C\left\{\left(1\vee \frac{q}{\sqrt{p}}\right)\sqrt{\frac{\log(p)}{n}} + \frac{q}{\sqrt{p}}\right\},~{j\in[p],}
$$
for some constant $C>0$, 
the estimator $\widehat{\gamma}_j$ provided in \eqref{gamma_j_lasso} satisfies the following estimation and prediction error bounds {uniformly for $j\in[p]$}:
$$
\|\widehat{\gamma}_j - \gamma_j\|_1 \lesssim_\pr \|D_{-j,-j}\left(\widehat{\gamma}_j - \gamma_j\right)\|_1  \lesssim_\pr A_n^2 s_j\lambda_j + \frac{s_j|R|_\infty^2}{\lambda_{j}} ,
$$
$$
\|F_\text{dc} X_{\cdot,-j}\left(\widehat{\gamma}_j - \gamma_j\right)\|_2 \lesssim_\pr \sqrt{ns_j}\left(A_n\lambda_j + |R|_\infty \right) ,
$$
where $R = F_\text{dc}X - E$ and $A_n = |R|_\infty \vee 1$.

\end{proposition}

To reach a concise theoretical rate result for the subsequent analysis, we additionally assume that $q \lesssim \sqrt{\log(p)}$ and obtain the following corollary.

\begin{corollary}\label{gammaj_error2}
Under Assumptions \ref{con_E_H}-\ref{con_E_independence}, if $q \lesssim \sqrt{\log(p)}$, $n \lesssim p$, $q\log(n)\sqrt{\log(p)} \lesssim \sqrt{n}$ and $\lambda_j \asymp \sqrt{\log(p) / n}$ uniformly for $j\in[p]$, we have
$$
\max_{j\in[p]} \| \widehat{\gamma}_j - \gamma_j \|_1  \lesssim_\pr \max_{j\in[p]} \|D_{-j,-j}\left(\widehat{\gamma}_j - \gamma_j\right)\|_1  \lesssim_\pr q^2\log^2(n)s_\Omega\sqrt{\frac{\log(p)}{n}},
$$
$$
\max_{j\in[p]} \|F_\text{dc} X_{\cdot,-j}\left(\widehat{\gamma}_j - \gamma_j\right)\|_2 \lesssim_\pr q\log(n)\sqrt{s_\Omega\log(p)}.
$$
\end{corollary}

\begin{remark}\label{remark:lambdaj}
Though a larger order of $\lambda_j \asymp q\log(n)\sqrt{\log(p) / n}$ produces faster estimation rates for $\gamma_j$'s according to Proposition \ref{gammaj_error}, 
we choose $\lambda_j \asymp \sqrt{\log(p) / n}$ because it leads to weaker conditions for the following theorems in order to control the second-order biases {of the debiased estimators}.
\end{remark}

\begin{remark}\label{remark:gamma_j}
It is also worthwhile to note that, though there is a similar intermediate result in \citet{guo2022}, the intuition and technical details are substantially different. 
In addition, to ensure the statistical independence in the technical analysis of \citet{guo2022}, node-wise Lasso procedures are performed based on separately constructed Trim transformations $F_{\text{T}}^{(j)}$, analog to \eqref{F_trim} while replacing $X$ by $X_{\cdot,-j}$. 
Hence, SVD will be applied $p$ times in their procedure for large-scale problems, which is computationally highly inefficient. 
In contrast, our decorrelating function is invariant for all node-wise Lasso steps and the technical dependence issue is solved by introducing a novel leave-one-out decorrelating function in Lemma 10 of the supplement.
\end{remark}

Based on the above error bounds in Propositions \ref{factor_estimation_bound} and \ref{gammaj_error}, we next establish the asymptotic normality result for the debiased estimator $\overline{\beta}_j$.
\begin{theorem}\label{asymp_normal}
	Under Assumptions \ref{con_E_H}-\ref{con_initial}, if $n \lesssim p$, $q \lesssim \sqrt{\log(p)}$, $\lambda_j \asymp \sqrt{\frac{\log(p)}{n}}$
	uniformly for $j\in[p]$, $ q s_0 = o\left\{\frac{ \sqrt{n} }{ \log(p) }\right\}$ and $ q^2 s_\Omega = o\left\{ \frac{n}{\log^2(n)\log^2(p)} \right\}$,
	then for any $j \in [p]$, we have
	$$
	\sqrt{n}\left( \overline{\beta}_j - \beta_j \right) \overset{d}{\rightarrow}N(0, \sigma_\xi^2 \omega_{j,j})
	\text{, and }
	\omega_{j,j} / \tau_j^2 \overset{\pr}{\rightarrow} 1.
	$$
\end{theorem}

The above theorem shows the asymptotic convergence of the debiased estimator under some mild sparsity assumptions and some appropriate initial estimators. Such asymptotic normality result is a key for the subsequent simultaneous inference.

\subsection{Comparisons of the Methods}\label{sec:compare}

This section studies the methodological and theoretical comparisons of the proposed debiased estimator $\overline{\beta}_j$ and the one in \citet{guo2022}. We start with the comparisons of their dependence structures and then explore and compare their testing efficiencies.

It is well known that, dependence structure is very crucial to ensure a provably valid multiple testing procedure. 
It may appear in the form of dependence among the $p$-values \citep[e.g.,][]{benjamini1995controlling,benjamini2001control,storey2004strong} or the correlations among the test statistics that are used to derive the asymptotic $p$-values \citep[e.g.,][]{liu2013,xia2015testing,cai2016,ma2021}.
As shown in the proof of Theorem \ref{asymp_normal}, the estimator $\overline{\beta}_j$ is asymptotically linear with the expansion $\sqrt{n}\left( \overline{\beta}_j - \beta_j \right) = \frac{\omega_{j,j}}{\sqrt{n}}\eta_j\trans \xi + \text{Re}_j$. 
Therefore, the dependence structure of the null statistics can be approximated through the correlations among $\eta_j\trans \xi$'s.
In contrast, the debiased estimator in \citet{guo2022} can be expressed as $\overline{\beta}_j^{\text{dd}} - \beta_j = \chi_j(X)\trans\xi + \text{Re}_j^{\text{dd}}$, where $\chi_j(X)\in\mathbb{R}^n$ is a function of $X$. 
It then follows that, $\chi_j(X)\trans\xi / \|\chi_j(X)\|_2 \sim \text{N}(0,1)$ if $\xi \sim \text{N}(0, \sigma_\xi^2 I_n)$.
While this way of establishing normality is useful for single hypothesis testing \citep[e.g.,][]{cai2021optimal,guo2021group},  it is in general challenging to perform multiple testing by directly incorporating such type of test statistics due to the complexity in characterizing the correlations among $\{ \chi_j(X)\trans\xi / \|\chi_j(X)\|_2\}_{j\in[p]}$.

On the other hand, by carefully exploring the dependence structure, the signal-noise ratio can be enhanced, which leads to a more powerful testing procedure \citep{fan2012estimating,fan2017estimation,du2021false}. 
According to Theorem 1 in \citet{guo2022}, the asymptotic variance of $\overline{\beta}_j^{\text{dd}}$ is given by $\sigma_\xi^2\omega_{j,j}\text{Tr}((F_\text{T}^{(j)})^4) / \text{Tr}^2((F_\text{T}^{(j)})^2)$.
By Cauchy-Schwarz inequality, it can be shown that such variance is larger than $\sigma_\xi^2\omega_{j,j} / n$, which is the asymptotic variance of the proposed estimator as presented in Theorem \ref{asymp_normal}.
Hence, the proposed decorrelating function \eqref{F_decorrelation} provides an improved signal-noise ratio asymptotically, which leads to a more efficient testing procedure subsequently.

\section{Simultaneous Inference} \label{sec:multi_testing}

We next develop the simultaneous testing procedure in Step 4 of Algorithm \ref{alg1}, so to identify the significant associations with the presence of the latent confounders. The theoretical properties in terms of both non-asymptotic and asymptotic false discovery control as well as the power analysis will be established.

\subsection{Data-Driven Multiple Testing Procedure}\label{sec:multiple-test}

Recall that, for $j\in [p]$, the test statistics are constructed by
$T_j = \frac{\sqrt{n}\overline{\beta}_j}{\widehat\sigma_\xi \tau_j}$ with $\tau_j = \sqrt{n}\| z_j \|_2^{-1}$.
It follows from Theorem \ref{asymp_normal} that a large value of $|T_j|$ indicates a strong evidence against the null $\mathcal{H}_{0,j}:\beta_j=0$. 
For the simultaneous inference problem \eqref{test}, we aim to choose an appropriate threshold $t$ for $|T_j|$ to control FDP and FDR.
Denote by $\mathcal{H}_0 = [p] \setminus S_0 = \{ j\in[p]: \beta_j=0 \}$ the index set containing all null hypotheses. Then the FDP at the threshold $t>0$ is given by
\begin{equation}\label{fdp}
	\text{FDP}(t) = \frac{\sum_{j\in \mathcal{H}_0}\mathbb{I}\left(|T_j|\geq t\right) }{R(t) \vee 1},
\end{equation}
where $R(t) = \sum_{j\in[p]}\mathbb{I}\left(|T_j|\geq t\right)$ is the total number of rejected hypotheses.
In order to detect as many significant hypotheses as possible while controlling the false discoveries, for a given level $\alpha\in(0,1)$, an ideal threshold can be determined by 
$
t_0=\inf \left\{t \geq 0: \text{FDP}(t) \leq \alpha\right\}.
$
However, the null set $\mathcal{H}_0$ is unknown, which makes the above procedure infeasible in practice.
As such, we first set to estimate the numerator term $\sum_{j\in \mathcal{H}_0}\mathbb{I}\left(|T_j|\geq t\right)$ of the FDP. Based on the asymptotic normality property in Theorem \ref{asymp_normal}, such quantity can be estimated by $|\mathcal{H}_0|G(t)$, where $G(t)=2(1 - \Phi(t))$ and $\Phi(\cdot)$ is the cumulative density function of standard Gaussian distribution. Due to the sparsity assumption on $\beta$, it can be further estimated by $pG(t)$.
Then the detailed simultaneous testing procedure is summarized in Algorithm \ref{alg2}, which serves as the last step of Algorithm \ref{alg1}.

\begin{algorithm}[htbp]

\caption{\label{alg2} Multiple Testing based on a Data-Driven Threshold.}
{Input:} test statistics \{$T_j$, $j\in[p]$\}, significance level $\alpha$.
\begin{enumerate}
\item[1.] \label{alg2:step1} 
	Calculate the threshold by
	\begin{equation}\label{t_hat}
		\widehat{t}=\inf \left\{0 \leq t \leq t_{p}: \frac{ pG(t) }{R(t) \vee 1} \leq \alpha\right\},
	\end{equation}
	where $t_p=\sqrt{2 \log (p)-2 \log\log(p)}$.
	If such $\widehat{t}$ in \eqref{t_hat} does not exist, set $\widehat{t}=\sqrt{2 \log(p)}$.

\item[2.] \label{alg2:step2} Reject  $\mathcal{H}_{0,j}$ if $j\in \widehat{S_0}$, where $\widehat{S_0} = \left\{j\in[p]: |T_j|\geq \widehat{t}\right\}$. 
Obtain the rejection sets for positive and negative signals respectively by
$$
\widehat{S}_{0+} = \left\{j\in\widehat{S_0}: T_j>0\right\},~\widehat{S}_{0-} = \left\{j\in\widehat{S_0}: T_j<0\right\}.
$$
\end{enumerate}
\end{algorithm}

\subsection{Non-Asymptotic FDR Control}\label{sec:finite}

Next we explore the error rate control of Algorithm \ref{alg2}.
We start with the finite-sample FDR control result and the asymptotic theories will be provided in Section \ref{sec:asy}. 

It is important to note that, the novel non-asymptotic analysis that will be established in Theorem \ref{nonasymptotic_fdr_control} is general and model-free. It can be employed in any scenarios as long as the test statistics $\{T_1,\ldots, T_p\}$ are provided and the threshold $\widehat{t}$ is determined by Algorithm \ref{alg2}. Therefore, such non-asymptotic FDR theory can be adopted in various simultaneous inference problems such as testing of high-dimensional covariance matrices, graphical models, generalized linear models \citep[e.g.,][]{xia2015testing,cai2016,javanmard2019,ma2021}, and is of independent interest. 

Define FDR=$\mathbb{E}\{\text{FDP}(\hat{t})\}$. The following theorem establishes the non-asymptotic FDR bound of Algorithm \ref{alg2}.

\begin{theorem}\label{nonasymptotic_fdr_control}
	For any $\epsilon>0$, we have
	$$
	\text{FDR} \leq \alpha\frac{|\mathcal{H}_0|}{p}(1+\epsilon) + \mathbb{P}\left(D_1 > \epsilon\right) + D_2, 
	$$
	where $D_1$ and $D_2$ are defined as
	$$
	D_1 = \sup_{j\in\mathcal{H}_0}\sup_{0\leq t \leq t_p}\left| \frac{ \mathbb{P}\left(|T_j|\geq t| T_{-j} \right) }{G(t)} - 1 \right|,~
	D_2 = \mathbb{P}\left\{ \sum_{j\in\mathcal{H}_0}\mathbb{I}\left( |T_j|\geq\sqrt{2\log(p)} \right) \geq 1 \right\},
	$$
	and $T_{-j} = (T_1,\ldots,T_{j-1},T_{j+1},\ldots,T_p)$.
\end{theorem}

We make a few explanations on Theorem \ref{nonasymptotic_fdr_control}. 
For any sufficiently small constant $\epsilon>0$, the FDR is upper bounded by the sum of three quantities, i.e., $\frac{|\mathcal{H}_0|}{p}\alpha(1+\epsilon)$, $\mathbb{P}\left(D_1 > \epsilon\right)$ and $D_2$.
These three terms in turn reflect the error rates corresponding to the weakly correlated case, highly correlated case, as well as the Gaussian deviated case.
First of all, if the null hypotheses are assumed to be independent or weakly correlated to each other, the bound of $\frac{|\mathcal{H}_0|}{p}\alpha(1+\epsilon)$ is well established in the literatures. 
Second, $\mathbb{P}\left(D_1 > \epsilon\right)$ reflects the probability of highly correlated scenarios and it quantifies the uniform relative deviation between the conditional tail probability $\mathbb{P}\left(|T_j|\geq t| T_{-j} \right)$
and the unconditional Gaussian tail probability $G(t)$, 
over all $0\leq t \leq t_p$ and all null indices $j\in\mathcal{H}_0$. 
Third, $D_2$ quantifies the excess rate of marginal Gaussian approximations, 
and it is negligible if the null distribution of $T_j$'s is well approximated by Gaussian distribution because $\mathbb{P}\left( \max_{j\in[p]}|Z_j| \geq \sqrt{2\log(p)} \right) = O\left( 1/\sqrt{\log(p)}\right)$ for standard Gaussian variables $\{Z_j,j\in[p]\}$.

\subsection{Asymptotic FDP and FDR Control}\label{sec:asy}

The above non-asymptotic result provides a guidance on the asymptotic analysis of this section. Particularly, the deviation term $D_1$ implies the key role of dependence in error rates control. 
By assuming a mild dependence condition, we obtain both asymptotic FDP and FDR control at a preassigned level $\alpha$ in Theorem \ref{fdp_fdr_control}.

\begin{assumption}\label{con_Omega_E}
For all $j \neq k$, $|\omega_{k,j}^0|\leq\theta$ for some constant $\theta \in (0,1)$, where $\omega_{k,j}^0 = \omega_{k,j}/(\omega_{k,k}\omega_{j,j})^{1/2}$.
\end{assumption}

\begin{theorem}\label{fdp_fdr_control}
	Under Assumptions \ref{con_E_H}-\ref{con_Omega_E}, if $n \lesssim p$, $q \lesssim \sqrt{\log(p)}$, $\{\log (p)\}^{7+\epsilon} \lesssim n$ for some small constant $\epsilon>0$, $\lambda_j \asymp \sqrt{\frac{\log(p)}{n}}$
	uniformly for $j\in[p]$, $ q s_0 = o\left\{\frac{\sqrt{n}}{\log^{3/2}(p)}\right\}$ and $ q^2 s_\Omega = o\left\{ \frac{n}{\log^2(n)\log^{3}(p)} \right\}$, we have
	$$
	\text{FDP}\left(\hat{t}\right) \leq \alpha + o_\pr(1), \text{ and } \limsup_{n\rightarrow \infty}\text{FDR}\leq \alpha.
	$$
\end{theorem}

Assumption \ref{con_Omega_E} is mild as it only excludes the cases with nearly perfect dependence.
Note that, a key step for the proof of Theorem \ref{fdp_fdr_control} is the Cramer-type Gaussian approximation that estimates $\sum_{j\in \mathcal{H}_0}\mathbb{I}\left(|T_j|\geq t\right)$ by $|\mathcal{H}_0|G(t)$. 
A modified approximation result is established in Lemma 7 of the supplement, and it improves the polynomial rate condition in \citet{liu2013,cai2016} to an exponential rate. Therefore, it provides a new technical toolbox for the general multiple testing problems.

\subsection{Power Analysis}\label{sec:power}

Now we turn to the asymptotic power analysis and provide a sufficient condition under which all signals can be detected with correct signs.
Intuitively, it is impossible to separate 
the null hypothesis $\mathcal{H}_{0,j}$ and alternative hypothesis $\mathcal{H}_{1,j}$ if the absolute value of $\beta_j$ vanishes at a fast rate.
Hence, a minimal signal strength condition is imposed in the following theorem for the support recovery and the asymptotic power analysis.

Define the index sets for the positive and negative signals by
	$S_{0+} = \left\{j\in[p]: \beta_j>0\right\}$ and $S_{0-} = \left\{j\in[p]: \beta_j<0\right\}$, respectively. Define the average power by
\begin{equation}\label{power_def}
\text{Power} = \mathbb{E}\frac{|S_0\cap \widehat{S_0}|}{s_0} = \frac{1}{s_0}\sum_{j\in S_0}\mathbb{P}\left(|T_j|\geq \widehat t\right).
\end{equation}
\begin{theorem}\label{power}
	Under Assumptions \ref{con_E_H}-\ref{con_initial}, if $n \lesssim p$, $q \lesssim \sqrt{\log(p)}$, $\{\log (p)\}^5 \lesssim n$, $\lambda_j \asymp \sqrt{\frac{\log(p)}{n}}$
	uniformly for $j\in[p]$, $ q s_0 = o\left\{ \sqrt{\frac{n}{\log(p)}} \right\}$, $ q^2 s_\Omega = o\left\{ \frac{n}{\log^2(n)\log(p)} \right\}$, and the minimal signal strength satisfies that
	$$
	\min_{j\in S_{0}}\frac{|\beta_j|}{\sigma_\xi\sqrt{\omega_{j,j}}} \geq (2+\epsilon)\sqrt{\frac{2\log(p)}{n}}
	$$ 
	for some small constant $\epsilon>0$, then we have
	$$\lim_{n\rightarrow \infty}\mathbb{P}\left( S_{0+} \subseteq \widehat{S}_{0+},  S_{0-} \subseteq \widehat{S}_{0-} \right) = 1.$$
	Consequently, it yields that Power $\rightarrow 1$ as $n\rightarrow \infty$.
\end{theorem}

Theorem \ref{power} states that all the signals can be detected with correct signs under a mild minimal signal strength condition and some sparsity assumptions on {$q$}, $s_0$ and $s_\Omega$ that are less restricted than those in Theorems \ref{asymp_normal} and \ref{fdp_fdr_control}. As a consequence, the multiple testing procedure proposed in Algorithm \ref{alg2} enjoys full power asymptotically. 

\section{Simulation Studies}\label{sec:simu}

In this section, we present the numerical performance of the proposed procedure in Algorithm \ref{alg1}.
Recall that, the initial estimator $\widehat{\beta}$ in \eqref{beta_lasso} can be obtained through either the decorrelating function $F_\text{dc}$ in \eqref{F_decorrelation} or the Trim function $F_\text{T}$ in \eqref{F_trim}. 
The two corresponding testing procedures are denoted by \texttt{Decorrelate \& Debias-dc} and \texttt{Decorrelate \& Debias-T}, respectively.
This section compares these two procedures with two competing methods that incorporate their statistics into our Algorithm \ref{alg2}: the approach in \citet{guo2022} (denoted by \texttt{Doubly Debias}), and the approach in \citet{zhang2014,van2014} (denoted by \texttt{Standard Debias}) that ignores the confounding effects.

\subsection{Data Generation and Implementation Details}\label{subsec:simu_data}

We generate data from the confounded linear regression model \eqref{model_confounder2} through the following mechanisms. First, the entries of the confounders $H$ and the noise $\xi$ are independently generated from $\text{N}(0,1)$, the loading $\Psi$ is set to have independent $\text{Uniform}(-2,2)$ entries, and the confounding coefficient vector $\phi$ is set to have independent $\text{N}(\mu,1)$ entries.
Next, three structures for $\Omega_E = (\omega_{j,k})$ are considered: identity, Erd\"os-R\'enyi random graph, and the banded graph; the detailed description for the graph construction is presented in Section E.1 of the supplement.
Finally, we randomly locate $s_0$ signals $\{\beta_{j_k}:k=1,\cdots,s_0\}$ with magnitude $|\beta_{j_k}| = 1.2^{-\nu}\left(8\omega_{j_k,j_k}\log (p) / n \right)^{1/2}$, and the sign of $\beta_{j_k}$ is drawn uniformly and randomly from $\{-1,1\}$.
In all simulation settings, we set $\mu = \nu = 3$.

Next, we describe the implementation details of the proposed methods as well as the competing approaches.
For the decorrelating function $F_{\text{dc}}$ in \eqref{F_decorrelation}, we employ the eigenvalue ratio method \citep{ahn2013} to estimate $q$. 
Specifically, let $\widehat{q}=\arg\max_{1\leq k\leq k_{\text{max}}}\frac{\Lambda_k\left( X\right)}{\Lambda_{k+1}\left( X \right)}$, and we set $k_{\text{max}}=20$.
For the Trim function in \eqref{F_trim}, the tuning parameter $\rho$  is set to be 0.3.
The details for the regularization parameter selection are presented in Sections E.2 and E.3 of the supplement.
To be specific, a calibrated error variance estimator that accommodates to the cross-validated $\lambda$ in procedure \eqref{beta_lasso} is proposed in Section E.2.
Additionally, we develop a new hyperparameter-free method in Section E.3 to select parameters $\lambda_j,j\in[p]$ in the nodewise Lasso procedures \eqref{gamma_j_lasso}.
All simulation results are based on $256$ independent replications with significance level $\alpha = 0.1$.

\subsection{FDR and Power Comparisons}\label{subsec:simu_fdr}

\begin{figure}[t]
	\centering
	\captionsetup{font=small}
    \includegraphics[scale=0.42]{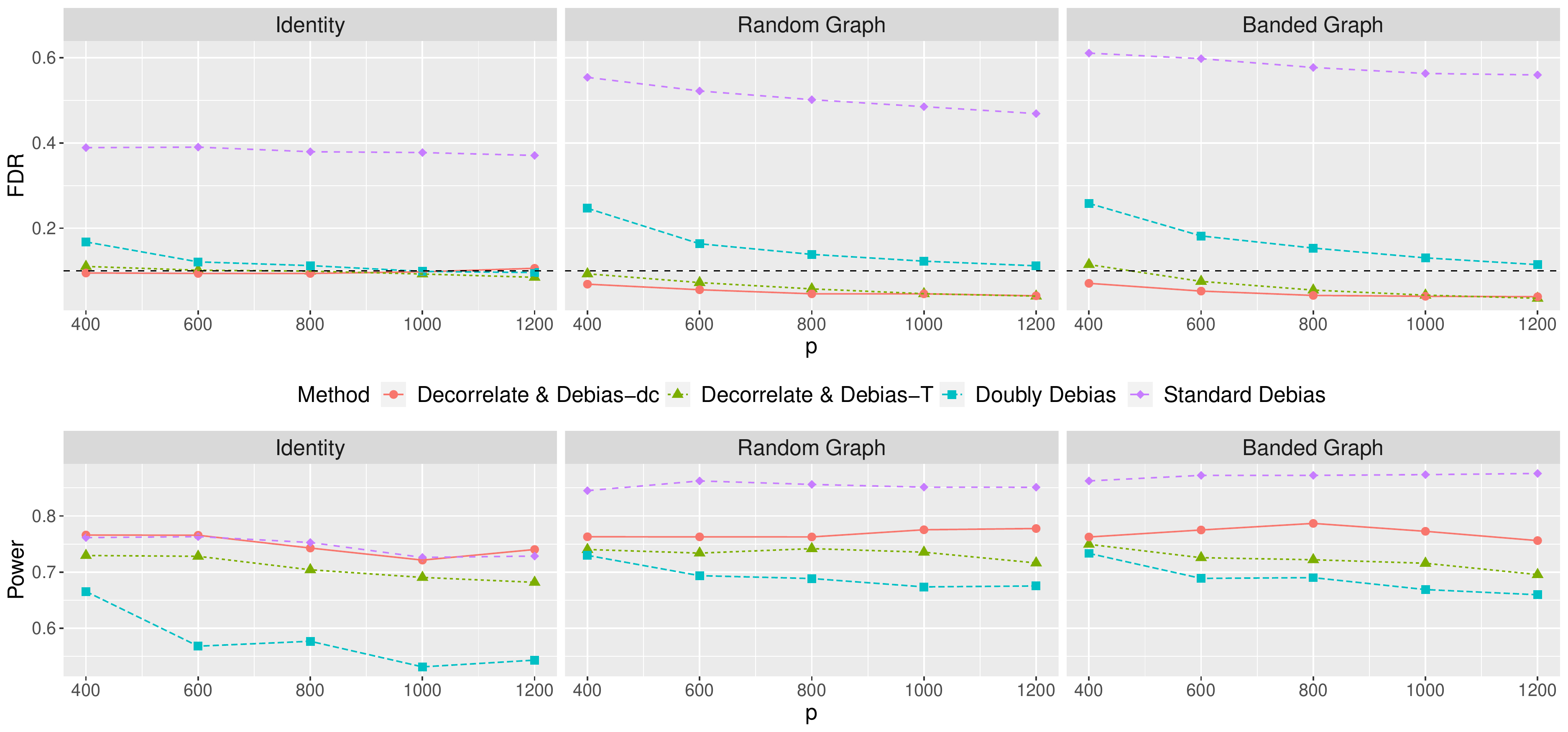}
    \caption{Empirical FDR and power comparisons with varying dimension $p$ and different graph structures; $(n, s_0, q) = (600, 30, 5)$ and $\alpha=0.1$.}
    \label{fig:p}
\end{figure}

To evaluate the performance of methods across different dimensions $p$, we fix $(n, s_0, q) = (600, 30, 5)$, and vary $p$ from 400 to 1200 in increments of 200.
The averaged FDP in \eqref{fdp} and averaged empirical power in \eqref{power_def} are reported in Figure \ref{fig:p}.
The top panel of Figure \ref{fig:p} shows that, the two proposed \texttt{Decorrelate \& Debias} methods have empirical FDR well under control across all settings.
In comparison, the \texttt{Standard Debias} method fails to achieve valid FDR control;
the \texttt{Doubly Debias} method exhibits unsatisfactory FDR performance in low and moderate dimensions for all graphs and it suffers from FDR inflation for complex graph structures across all dimensions.
This performance agrees with the discussions in Section \ref{sec:compare} that
 the validity for the adoption of \texttt{Doubly Debias} estimator in multiple testing problem warrants further theoretical explorations.  
On the other hand, the bottom panel of Figure \ref{fig:p} shows that, both of the proposed \texttt{Decorrelate \& Debias} methods exhibit superior power performance across all settings.
We also observe different performance of the two proposed methods in terms of FDR or power. 
This discrepancy may arise from the impact of finite sample experiments or from the subtle effect of the initial estimators on the second-order biases \citep{javanmard2018debiasing}.

\begin{figure}[t]
	\centering
	\captionsetup{font=small}
    \includegraphics[scale=0.42]{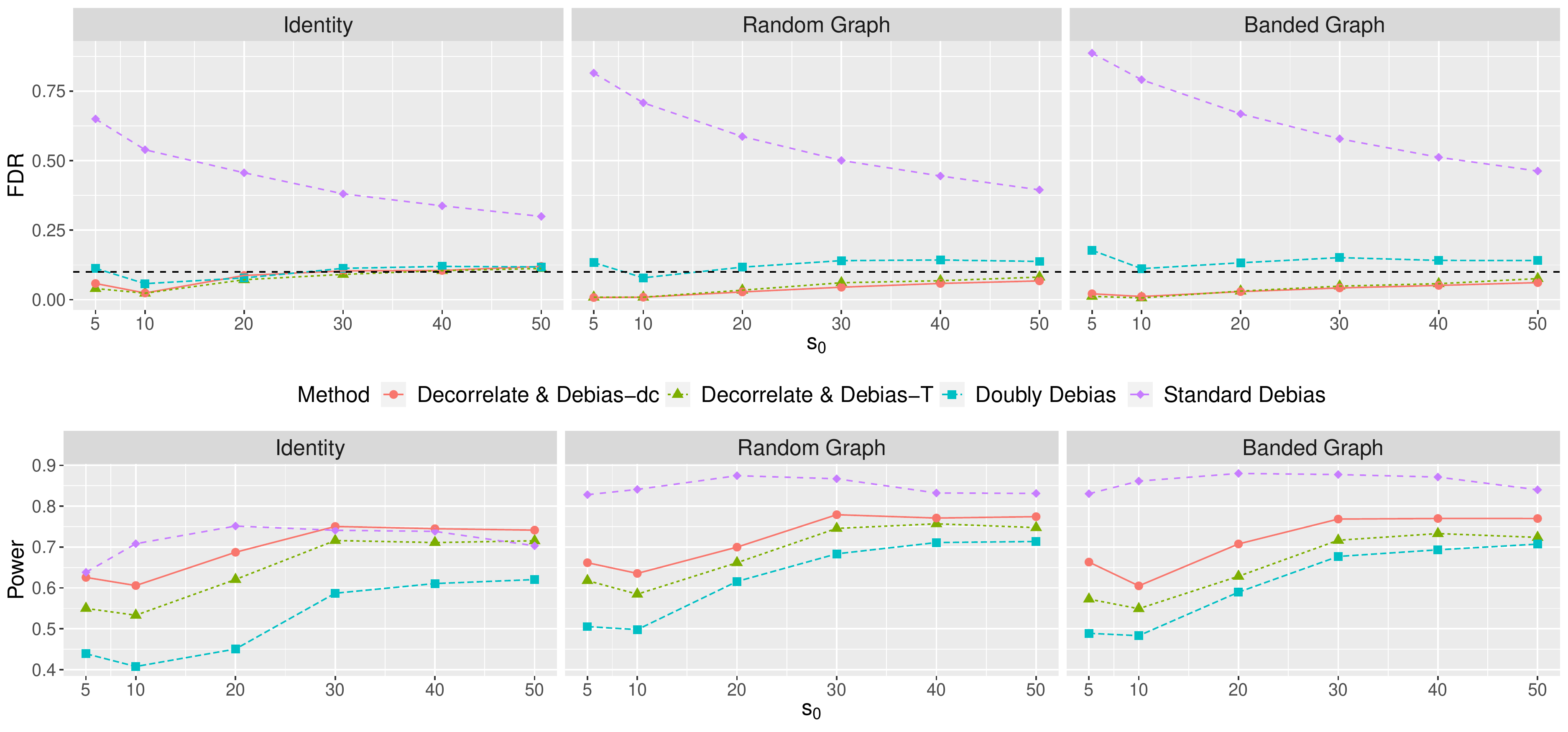}
    \caption{Empirical FDR and power comparisons with varying sparsity level $s_0$  and different graph structures; $(n, p, q) = (600, 800, 5)$ and $\alpha=0.1$.}
    \label{fig:s0}
\end{figure}

To evaluate the effect of sparsity levels, we next fix $(n, p, q) = (600, 800, 5)$, and vary $s_0$ from 10 to 50 in increments of 10.
Besides, we include a setting with $s_0 = 5$ to explore the performance when $s_0 < 1 / \alpha$.
The results are summarized in Figure \ref{fig:s0}.
The top panel shows that both of the proposed methods achieve FDR control across all sparsity levels. Similarly as the   results in Figure \ref{fig:p}, \texttt{Standard Debias} cannot attain FDR control while \texttt{Doubly Debias} has some FDR inflations for complex graph structures.
Again, the bottom panel of Figure \ref{fig:s0} illustrates the power advantage of the two proposed methods compared to \texttt{Doubly Debias} method.

Finally, we investigate the impact of misspecified number of confounders $q$ on the proposed methods.
The experimental details and results are presented in Section E.4 of the supplement due to space limitations.
Figure E.2 indicates that an overestimation of $q$ has little impact on the performance of the proposed methods in terms of both FDR and power; this is consistent with the findings in \citet{fan2013}.
Hence,  in practice, a slightly larger estimation of $q$ is preferable in the construction of the decorrelating function \eqref{F_decorrelation}.

\section{Real Data Analysis}\label{sec:real_data}

In this section, we study the associations between the drug resistance and the genotype mutations on the Human Immunodeficiency Virus Type 1 (HIV-1) datasets in \citet{rhee2006genotypic}.
The datasets contain three distinct drug classes: protease inhibitors (PI), nucleoside reverse transcriptase (RT) inhibitors (NRTI), and non-nucleoside RT inhibitors (NNRTI).
For each drug class, the data comprise the HIV-1 protease/RT mutations, along with the corresponding resistance measurements for several different drugs.
Our objective is to simultaneously identify positions of mutations that are associated  with the drug resistance.
To assess our results, the selected positions are compared with sets of non-polymorphic treatment-selected mutations (TSM) \citep{rhee2005hiv}.
Similarly as done in \cite{barber2015controlling}, the TSM sets provide an approximation to the ground truth and can be employed to evaluate different methods.
Both genotype-phenotype and TSM datasets can be accessed at \url{https://hivdb.stanford.edu/pages/published_analysis/genophenoPNAS2006/}.

The data processing steps are briefly described as follows.
In line with \citet{barber2015controlling}, the design matrices for each drug class are constructed based on the presence/absence of mutations, and the responses are the logarithmic drug resistance measurements.
In each study, the samples with missing responses are excluded, and the duplicate and {zero columns} are removed from the design matrices.
Subsequently, all variables are scaled to have zero means and unit variances.
Note that PI class is not suitable for our analysis due to the low-dimensionality of its resulting designs. 
Therefore, we focus on the NRTI and NNRTI drug classes, comprising six and three drugs, respectively.
Figure E.3 in the supplement presents the singular value distributions as well as the sample sizes and dimensions of the designs in the target datasets; see Section E.5 for more details including the determination of  the numbers of confounders.

\begin{figure}[t]
	\centering
	\captionsetup{font=small}
    \includegraphics[scale=0.42]{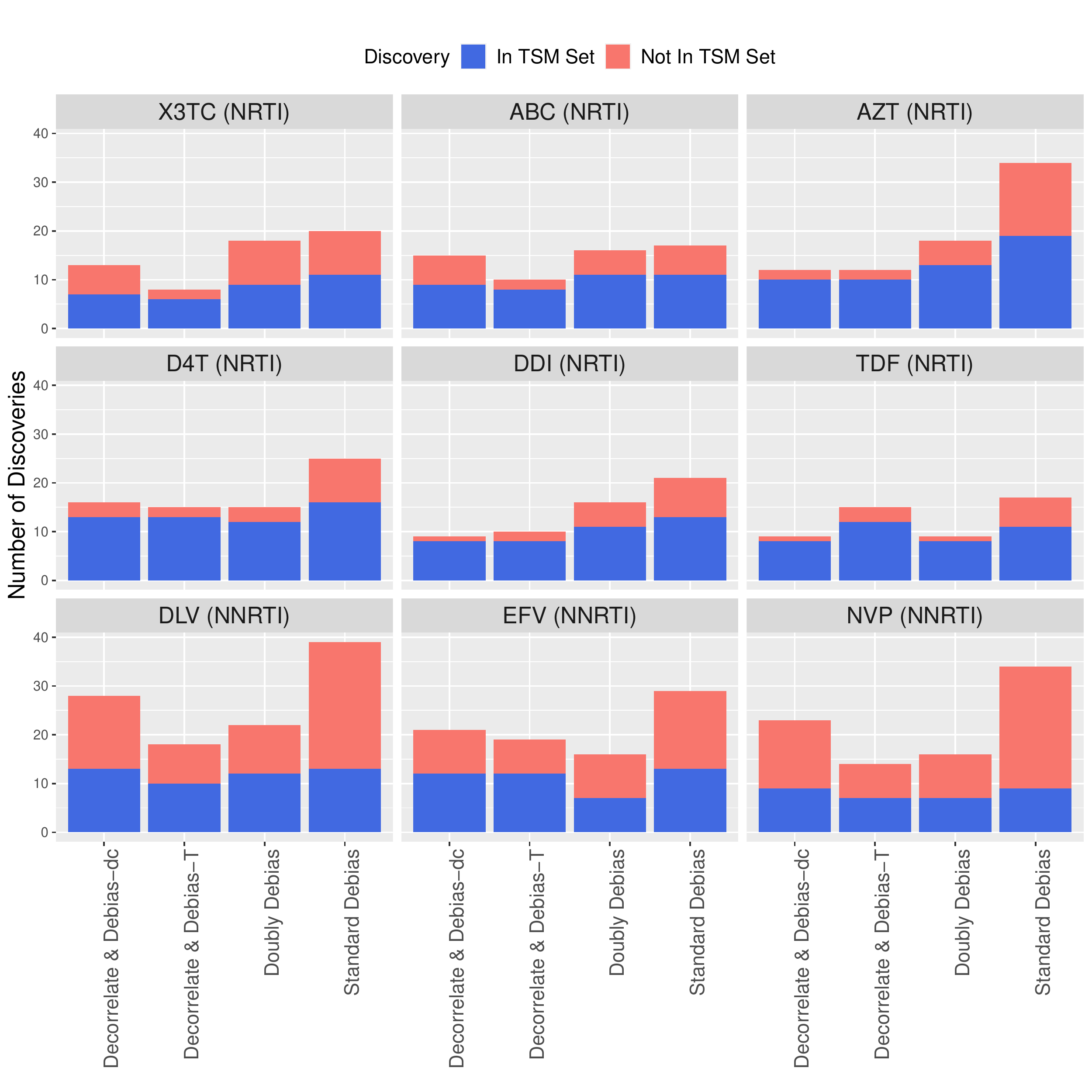}
    \caption{Numbers of selected positions of mutations on the HIV-1 datasets by applying different methods. The bottom blue indicates the number of discoveries that appear in the TSM sets; the {top} red indicates the number of discoveries that do not appear in the TSM sets.}
    \label{fig:res_plot}
\end{figure}

We now apply the proposed \texttt{Decorrelate \& Debias-dc} and \texttt{Decorrelate \& Debias-T} as well as the two competing methods to the processed datasets, and perform the simultaneous association analysis with the error rate $\alpha=0.1$.
To present the findings, multiple types of mutations occurring at the same position are treated indistinctly \citep{barber2015controlling}.
The numbers of selected positions for different methods are displayed in Figure \ref{fig:res_plot}, where the bottom blue/{top} red boxes respectively indicate the counts of discoveries that are consistent/inconsistent with the TSM sets.
Within the NRTI class, the results obtained by the proposed methods show better agreement with TSM compared to other methods and in the meanwhile demonstrate power competitiveness.
For the alternative methods, \texttt{Doubly Debias} shows comparable performance in many cases, with exceptions including the drugs X3TC and DDI where there are an excessive number of discoveries that are inconsistent with TSM;
\texttt{Standard Debias} discovers many positions that are not included in TSM.
For the NNRTI class, the overall performance of the methods is similar to that in the NRTI class. 
However, the discovery inconsistency with TSM becomes more apparent in NNRTI, as also observed in \citet{barber2015controlling,fithian2022conditional}. 
Such phenomenon suggests that the TSM sets initially reported by \citet{rhee2005hiv} may deserve potential expansions and further scientific explorations.

\bibliographystyle{apalike}
\bibliography{ref2}

\begin{thebibliography}{}

\bibitem[Ahn and Horenstein, 2013]{ahn2013}
Ahn, S.~C. and Horenstein, A.~R. (2013).
\newblock Eigenvalue ratio test for the number of factors.
\newblock {\em Econometrica}, 81(3):1203--1227.

\bibitem[Bai and Ng, 2002]{bai2002}
Bai, J. and Ng, S. (2002).
\newblock Determining the number of factors in approximate factor models.
\newblock {\em Econometrica}, 70(1):191--221.

\bibitem[Barber and Cand{\`e}s, 2015]{barber2015controlling}
Barber, R.~F. and Cand{\`e}s, E.~J. (2015).
\newblock Controlling the false discovery rate via knockoffs.
\newblock {\em Ann. Stat.}, 43(5):2055--2085.

\bibitem[Benjamini and Hochberg, 1995]{benjamini1995controlling}
Benjamini, Y. and Hochberg, Y. (1995).
\newblock Controlling the false discovery rate: a practical and powerful
  approach to multiple testing.
\newblock {\em J. R. Stat. Soc. B}, 57(1):289--300.

\bibitem[Benjamini and Yekutieli, 2001]{benjamini2001control}
Benjamini, Y. and Yekutieli, D. (2001).
\newblock The control of the false discovery rate in multiple testing under
  dependency.
\newblock {\em Ann. Stat.}, 29(4):1165--1188.

\bibitem[Bickel et~al., 2009]{bickel2009}
Bickel, P.~J., Ritov, Y., and Tsybakov, A.~B. (2009).
\newblock Simultaneous analysis of lasso and dantzig selector.
\newblock {\em Ann. Stat.}, 37(4):1705--1732.

\bibitem[Bing et~al., 2022a]{bing2022inference}
Bing, X., Cheng, W., Feng, H., and Ning, Y. (2022a).
\newblock Inference in high-dimensional multivariate response regression with
  hidden variables.
\newblock {\em arXiv preprint arXiv:2201.08003}.

\bibitem[Bing et~al., 2022b]{bing2022estimation}
Bing, X., Ning, Y., and Xu, Y. (2022b).
\newblock {Adaptive estimation in multivariate response regression with hidden
  variables}.
\newblock {\em Ann. Stat.}, 50(2):640--672.

\bibitem[Cai et~al., 2021]{cai2021optimal}
Cai, T., Cai, T.~T., and Guo, Z. (2021).
\newblock Optimal statistical inference for individualized treatment effects in
  high-dimensional models.
\newblock {\em J. R. Stat. Soc. B}, 83(4):669--719.

\bibitem[Cai and Liu, 2016]{cai2016}
Cai, T.~T. and Liu, W. (2016).
\newblock Large-scale multiple testing of correlations.
\newblock {\em J. Am. Stat. Assoc.}, 111(513):229--240.

\bibitem[{\'C}evid et~al., 2020]{cevid2020}
{\'C}evid, D., B{\"u}hlmann, P., and Meinshausen, N. (2020).
\newblock Spectral deconfounding via perturbed sparse linear models.
\newblock {\em J. Mach. Learn. Res.}, 21(232):1--41.

\bibitem[Chernozhukov et~al., 2017]{chernozhukov2017}
Chernozhukov, V., Hansen, C., and Liao, Y. (2017).
\newblock A lava attack on the recovery of sums of dense and sparse signals.
\newblock {\em Ann. Stat.}, 45(1):39--76.

\bibitem[Dezeure et~al., 2017]{dezeure2017}
Dezeure, R., B{\"u}hlmann, P., and Zhang, C.-H. (2017).
\newblock High-dimensional simultaneous inference with the bootstrap.
\newblock {\em Test}, 26(4):685--719.

\bibitem[Du et~al., 2023]{du2021false}
Du, L., Guo, X., Sun, W., and Zou, C. (2023).
\newblock False discovery rate control under general dependence by symmetrized
  data aggregation.
\newblock {\em J. Am. Stat. Assoc.}, 118(541):607--621.

\bibitem[Efron, 2007]{efron2007correlation}
Efron, B. (2007).
\newblock Correlation and large-scale simultaneous significance testing.
\newblock {\em J. Am. Stat. Assoc.}, 102(477):93--103.

\bibitem[Fan et~al., 2022]{fan2022}
Fan, J., Guo, J., and Zheng, S. (2022).
\newblock Estimating number of factors by adjusted eigenvalues thresholding.
\newblock {\em J. Am. Stat. Assoc.}, 117(538):852--861.

\bibitem[Fan and Han, 2017]{fan2017estimation}
Fan, J. and Han, X. (2017).
\newblock Estimation of the false discovery proportion with unknown dependence.
\newblock {\em J. R. Stat. Soc. B}, 79(4):1143--1164.

\bibitem[Fan et~al., 2012]{fan2012estimating}
Fan, J., Han, X., and Gu, W. (2012).
\newblock Estimating false discovery proportion under arbitrary covariance
  dependence.
\newblock {\em J. Am. Stat. Assoc.}, 107(499):1019--1035.

\bibitem[Fan et~al., 2020]{fankewang2020}
Fan, J., Ke, Y., and Wang, K. (2020).
\newblock Factor-adjusted regularized model selection.
\newblock {\em J. Econom.}, 216(1):71--85.

\bibitem[Fan and Li, 2001]{fan2001}
Fan, J. and Li, R. (2001).
\newblock Variable selection via nonconcave penalized likelihood and its oracle
  properties.
\newblock {\em J. Am. Stat. Assoc.}, 96(456):1348--1360.

\bibitem[Fan et~al., 2013]{fan2013}
Fan, J., Liao, Y., and Mincheva, M. (2013).
\newblock Large covariance estimation by thresholding principal orthogonal
  complements.
\newblock {\em J. R. Stat. Soc. B}, 75(4):603--680.

\bibitem[Fan et~al., 2018]{fan2018}
Fan, J., Liu, H., and Wang, W. (2018).
\newblock Large covariance estimation through elliptical factor models.
\newblock {\em Ann. Stat.}, 46(4):1383--1414.

\bibitem[Fithian and Lei, 2022]{fithian2022conditional}
Fithian, W. and Lei, L. (2022).
\newblock Conditional calibration for false discovery rate control under
  dependence.
\newblock {\em Ann. Stat.}, 50(6):3091--3118.

\bibitem[Guo et~al., 2022]{guo2022}
Guo, Z., {\'C}evid, D., and B{\"u}hlmann, P. (2022).
\newblock Doubly debiased lasso: High-dimensional inference under hidden
  confounding.
\newblock {\em Ann. Stat.}, 50(3):1320--1347.

\bibitem[Guo et~al., 2021]{guo2021group}
Guo, Z., Renaux, C., B{\"u}hlmann, P., and Cai, T.~T. (2021).
\newblock Group inference in high dimensions with applications to hierarchical
  testing.
\newblock {\em Electron. J. Stat.}, 15(2):6633--6676.

\bibitem[Hsu et~al., 2012]{hsu2012reducing}
Hsu, F.-H., Serpedin, E., Hsiao, T.-H., Bishop, A.~J., Dougherty, E.~R., and
  Chen, Y. (2012).
\newblock Reducing confounding and suppression effects in tcga data: an
  integrated analysis of chemotherapy response in ovarian cancer.
\newblock {\em BMC Genomics}, 13(6):1--15.

\bibitem[Javanmard and Javadi, 2019]{javanmard2019}
Javanmard, A. and Javadi, H. (2019).
\newblock False discovery rate control via debiased lasso.
\newblock {\em Electron. J. Stat.}, 13(1):1212--1253.

\bibitem[Javanmard and Montanari, 2014]{javanmard2014}
Javanmard, A. and Montanari, A. (2014).
\newblock Confidence intervals and hypothesis testing for high-dimensional
  regression.
\newblock {\em J. Mach. Learn. Res.}, 15(1):2869--2909.

\bibitem[Javanmard and Montanari, 2018]{javanmard2018debiasing}
Javanmard, A. and Montanari, A. (2018).
\newblock Debiasing the lasso: Optimal sample size for gaussian designs.
\newblock {\em Ann. Stat.}, 46(6A):2593--2622.

\bibitem[Lahiri, 2021]{lahiri2021}
Lahiri, S.~N. (2021).
\newblock Necessary and sufficient conditions for variable selection
  consistency of the lasso in high dimensions.
\newblock {\em Ann. Stat.}, 49(2):820--844.

\bibitem[Leek et~al., 2010]{leek2010tackling}
Leek, J.~T., Scharpf, R.~B., Bravo, H.~C., Simcha, D., Langmead, B., Johnson,
  W.~E., Geman, D., Baggerly, K., and Irizarry, R.~A. (2010).
\newblock Tackling the widespread and critical impact of batch effects in
  high-throughput data.
\newblock {\em Nat. Rev. Genet.}, 11(10):733--739.

\bibitem[Leek and Storey, 2007]{leek2007capturing}
Leek, J.~T. and Storey, J.~D. (2007).
\newblock Capturing heterogeneity in gene expression studies by surrogate
  variable analysis.
\newblock {\em PLoS Genet.}, 3(9):e161.

\bibitem[Leek and Storey, 2008]{leek2008general}
Leek, J.~T. and Storey, J.~D. (2008).
\newblock A general framework for multiple testing dependence.
\newblock {\em Proc. Natl. Acad. Sci.}, 105(48):18718--18723.

\bibitem[Liu, 2013]{liu2013}
Liu, W. (2013).
\newblock Gaussian graphical model estimation with false discovery rate
  control.
\newblock {\em Ann. Stat.}, 41(6):2948--2978.

\bibitem[Liu and Luo, 2014]{liu2014}
Liu, W. and Luo, S. (2014).
\newblock Hypothesis testing for high-dimensional regression models.
\newblock {\em Technical Report}.

\bibitem[Ma et~al., 2021]{ma2021}
Ma, R., Cai, T.~T., and Li, H. (2021).
\newblock Global and simultaneous hypothesis testing for high-dimensional
  logistic regression models.
\newblock {\em J. Am. Stat. Assoc.}, 116(534):984--998.

\bibitem[Meinshausen and B{\"u}hlmann, 2006]{meinshausen2006}
Meinshausen, N. and B{\"u}hlmann, P. (2006).
\newblock High-dimensional graphs and variable selection with the lasso.
\newblock {\em Ann. Stat.}, 34(3):1436--1462.

\bibitem[Negahban et~al., 2012]{negahban2012}
Negahban, S.~N., Ravikumar, P., Wainwright, M.~J., and Yu, B. (2012).
\newblock A unified framework for high-dimensional analysis of $ m $-estimators
  with decomposable regularizers.
\newblock {\em Stat. Sci.}, 27(4):538--557.

\bibitem[Reid et~al., 2016]{reid2016study}
Reid, S., Tibshirani, R., and Friedman, J. (2016).
\newblock A study of error variance estimation in lasso regression.
\newblock {\em Stat. Sin.}, 26:35--67.

\bibitem[Rhee et~al., 2005]{rhee2005hiv}
Rhee, S.-Y., Fessel, W.~J., Zolopa, A.~R., Hurley, L., Liu, T., Taylor, J.,
  Nguyen, D.~P., Slome, S., Klein, D., Horberg, M., et~al. (2005).
\newblock Hiv-1 protease and reverse-transcriptase mutations: correlations with
  antiretroviral therapy in subtype b isolates and implications for
  drug-resistance surveillance.
\newblock {\em J. Infect. Dis.}, 192(3):456--465.

\bibitem[Rhee et~al., 2006]{rhee2006genotypic}
Rhee, S.-Y., Taylor, J., Wadhera, G., Ben-Hur, A., Brutlag, D.~L., and Shafer,
  R.~W. (2006).
\newblock Genotypic predictors of human immunodeficiency virus type 1 drug
  resistance.
\newblock {\em Proc. Natl. Acad. Sci.}, 103(46):17355--17360.

\bibitem[Schwartz and Coull, 2003]{schwartz2003control}
Schwartz, J. and Coull, B.~A. (2003).
\newblock Control for confounding in the presence of measurement error in
  hierarchical models.
\newblock {\em Biostatistics}, 4(4):539--553.

\bibitem[Sheppard et~al., 2012]{sheppard2012confounding}
Sheppard, L., Burnett, R.~T., Szpiro, A.~A., Kim, S.-Y., Jerrett, M., Pope,
  C.~A., and Brunekreef, B. (2012).
\newblock Confounding and exposure measurement error in air pollution
  epidemiology.
\newblock {\em Air Qual. Atmos. Health}, 5(2):203--216.

\bibitem[Sila et~al., 2016]{sila2016women}
Sila, V., Gonzalez, A., and Hagendorff, J. (2016).
\newblock Women on board: Does boardroom gender diversity affect firm risk?
\newblock {\em J. Corp. Finance}, 36:26--53.

\bibitem[Stock and Watson, 2002]{stock2002}
Stock, J.~H. and Watson, M.~W. (2002).
\newblock Forecasting using principal components from a large number of
  predictors.
\newblock {\em J. Am. Stat. Assoc.}, 97(460):1167--1179.

\bibitem[Storey et~al., 2004]{storey2004strong}
Storey, J.~D., Taylor, J.~E., and Siegmund, D. (2004).
\newblock Strong control, conservative point estimation and simultaneous
  conservative consistency of false discovery rates: a unified approach.
\newblock {\em J. R. Stat. Soc. B}, 66(1):187--205.

\bibitem[Tibshirani, 1996]{tibshirani1996}
Tibshirani, R. (1996).
\newblock Regression shrinkage and selection via the lasso.
\newblock {\em J. R. Stat. Soc. B}, 58(1):267--288.

\bibitem[van~de Geer and B{\"u}hlmann, 2009]{van2009}
van~de Geer, S. and B{\"u}hlmann, P. (2009).
\newblock On the conditions used to prove oracle results for the lasso.
\newblock {\em Electron. J. Stat.}, 3:1360--1392.

\bibitem[van~de Geer et~al., 2014]{van2014}
van~de Geer, S., B{\"u}hlmann, P., Ritov, Y., and Dezeure, R. (2014).
\newblock On asymptotically optimal confidence regions and tests for
  high-dimensional models.
\newblock {\em Ann. Stat.}, 42(3):1166--1202.

\bibitem[Wang and Ramdas, 2022]{wang2022false}
Wang, R. and Ramdas, A. (2022).
\newblock False discovery rate control with e-values.
\newblock {\em J. R. Stat. Soc. B}, 84(3):822--852.

\bibitem[Xia et~al., 2015]{xia2015testing}
Xia, Y., Cai, T., and Cai, T.~T. (2015).
\newblock Testing differential networks with applications to the detection of
  gene-gene interactions.
\newblock {\em Biometrika}, 102(2):247--266.

\bibitem[Xia et~al., 2018]{xia2018}
Xia, Y., Cai, T., and Cai, T.~T. (2018).
\newblock Two-sample tests for high-dimensional linear regression with an
  application to detecting interactions.
\newblock {\em Stat. Sin.}, 28(1):63--92.

\bibitem[Zhang and Zhang, 2014]{zhang2014}
Zhang, C.-H. and Zhang, S.~S. (2014).
\newblock Confidence intervals for low dimensional parameters in high
  dimensional linear models.
\newblock {\em J. R. Stat. Soc. B}, 76(1):217--242.

\bibitem[Zhang and Cheng, 2017]{zhang2017}
Zhang, X. and Cheng, G. (2017).
\newblock Simultaneous inference for high-dimensional linear models.
\newblock {\em J. Am. Stat. Assoc.}, 112(518):757--768.

\bibitem[Zhao and Yu, 2006]{zhao2006}
Zhao, P. and Yu, B. (2006).
\newblock On model selection consistency of lasso.
\newblock {\em J. Mach. Learn. Res.}, 7:2541--2563.

\bibitem[Zou, 2006]{zou2006adaptive}
Zou, H. (2006).
\newblock The adaptive lasso and its oracle properties.
\newblock {\em J. Am. Stat. Assoc.}, 101(476):1418--1429.

\end{thebibliography}

\end{document}